\documentclass[aps,prl,reprint,footnotebib,amsfonts,amsmath,amssymb,superscriptaddress]{revtex4-2}

\usepackage{graphicx}
\usepackage{color}
\usepackage{tabularx} %

\usepackage[bookmarks=true,%
    colorlinks=true,%
    linkcolor=blue,%
    citecolor=blue,%
    filecolor=blue,%
    menucolor=blue,%
    urlcolor=blue,%
    pdfduplex=DuplexFlipLongEdge,%
    breaklinks=true]{hyperref}

\usepackage{latexsym,exscale,enumerate,comment, bbold}
\usepackage{amsthm,amscd}
\usepackage{subfigure}
\usepackage{slashed}
\usepackage{braket}
\usepackage[all,knot]{xy}
\usepackage{tikz}
\usepackage{upgreek}

\usetikzlibrary{decorations.markings}
\usetikzlibrary{decorations.pathreplacing}
\tikzstyle directed=[postaction={decorate,decoration={markings,
    mark=at position #1 with {\arrow{>}}}}]

\theoremstyle{plain}

\theoremstyle{definition}

\theoremstyle{definition}

\newcommand{\Tr}{{\rm Tr}}

\makeatletter
\let\oldhat\hat

\renewcommand{\hat}{\mathpalette\smart@hat}
\newcommand{\smart@hat}[2]{%
	\begingroup
	\setbox0=\hbox{$#1#2$}%
	\setbox2=\hbox{$#1M$}%
	\dimen0=\wd2
	\multiply\dimen0 by 6 \divide\dimen0 by 5
	\ifdim\wd0<\dimen0
	\oldhat{#2}%
	\else
	\widehat{#2}%
	\fi
	\endgroup
}

\let\oldtilde\tilde
\renewcommand{\tilde}{\mathpalette\smart@tilde}
\newcommand{\smart@tilde}[2]{%
	\begingroup
	\setbox0=\hbox{$#1#2$}%
	\setbox2=\hbox{$#1\Delta$}%
	\dimen0=\wd2
	\multiply\dimen0 by 6 \divide\dimen0 by 5
	\ifdim\wd0<\dimen0
	\oldtilde{#2}%
	\else
	\widetilde{#2}%
	\fi
	\endgroup
}
\makeatother

\let\epsilon=\varepsilon

\usepackage{bbm}

\def\1{\mathbbm{1}}%
\usepackage{bm}

\allowdisplaybreaks

\usepackage[normalem]{ulem}

\begin{document}
\title{Boundary criticality in two-dimensional correlated topological superconductors}

\author{Yang Ge}
\altaffiliation{These authors contributed equally.}
\affiliation{Department of Physics and Engineering Physics, Tulane University, New Orleans, Louisiana 70118, USA}

\author{Huan Jiang}
\altaffiliation{These authors contributed equally.}
\affiliation{Department of Physics and Engineering Physics, Tulane University, New Orleans, Louisiana 70118, USA}

\author{Hong Yao}
\email{yaohong@tsinghua.edu.cn}
\affiliation{Institute for Advanced Study, Tsinghua University, Beijing 100084, China}

\author{Shao-Kai Jian}
\email{sjian@tulane.edu}
\affiliation{Department of Physics and Engineering Physics, Tulane University, New Orleans, Louisiana 70118, USA}

\maketitle
\setcounter{tocdepth}{3}

{\bf 
The presence of a boundary enriches the nature of quantum phase transitions. 
However, the boundary critical phenomena in topological superconductors remain underexplored so far. 
Here, we investigate the boundary criticality in a two-dimensional correlated time-reversal-invariant topological superconductor tuned through a quantum phase transition into a trivial time-reversal-breaking superconductor. 
Using sign-problem-free determinant quantum Monte Carlo simulations, we chart the quantum phase diagram and reveal the boundary criticalities encompassing ordinary, special, and extraordinary transitions. 
Additionally, using renormalization group analysis, we compute the boundary critical exponent up to two loops. 
Remarkably, the simulations and two-loop renormalization group calculations consistently demonstrate that the presence of the boundary Majorana fermion at the special transition gives rise to a new type of boundary Gross-Neveu-Yukawa fixed point. 
We conclude with a discussion of possible experimental realizations in iron-based superconductors.
}

Boundary criticality has re-emerged as a vibrant area of research, revealing exotic quantum phenomena localized at surfaces and edges that enrich bulk universality classes~\cite{andrei2018boundary,cardy1984conformal,diehl1986field,mcavity1995conformal,diehl1997the}. 
Recent advances have uncovered a host of new boundary universality classes and scaling behaviors. In the bosonic setting, tremendous progress has been made on the $O(N)$ models: the discovery of the extraordinary-log transition for intermediate $N$ values has reshaped the landscape of surface criticality~\cite{metlitski2020boundary}, with conformal bootstrap~\cite{padayasi2021the} and Monte Carlo studies~\cite{toldin2021boundary,hu2021extraordinary-log,zou2022surface,zhang2022pott,sun2022villain,toldin2025universal} converging on its existence. 
More recently, the tricritical $O(N)$ model has provided a striking example in which boundary order parameters spontaneously order, thereby evading the Mermin-Wagner theorem in the context of boundary conformal field theory (BCFT)~\cite{sun2025boundary}.
 
Beyond purely bosonic models, there has been accelerating progress in fermionic boundary conformal field theories. 
Studies of Gross-Neveu and Gross-Neveu-Yukawa (GNY) models in the presence of boundaries have identified rich boundary phase structures, highlighting how fermionic degrees of freedom qualitatively alter boundary criticality~\cite{giombi2021fermions,herzog2023fermions,dipietro20203d,dipietro2023conformal,behan2021bootstrapping}. 
In particular, using $\epsilon$-expansion analyses and determinant quantum Monte Carlo simulations, the recent work~\cite{jiang2025boundary} has revealed the GNY boundary universality class in condensed matter systems.

These advances naturally motivate the exploration of boundary critical phenomena in symmetry-protected topological (SPT) phases~\cite{kitaev2009periodic,ryu2010topological,qi2011topological,hasan2010colloquium}, where symmetry-protected fermionic modes coexist with bulk critical excitations. 
Recent theoretical development has revealed novel boundary universality classes in such systems, including boundary GNY and special Berezinskii-Kosterlitz-Thouless transitions, driven by the interplay between boundary fermions and bulk critical order parameters in topological insulators and superconductors~\cite{shen2024new,ge2025tibcft}. 
Notice that the boundary critical phenomena in related topological systems have also been studied in Refs.~\onlinecite{scaffidi2017gapless,zhang2017unconventional,wu2020boundary,xu2020topological,verresen2021gapless,ma2022edge,yu2022conformal,ji2023boundary,yu2024universal,toldin2025extraordinary,liu2025edge}.

The boundary GNY universality class features a strongly coupled system between $d$-dimensional Ising BCFT and $(d-1)$-dimensional relativistic fermion theory. 
Time-reversal invariant (TRI) topological superconductors (TSCs) offer a natural platform for such boundary criticality. 
They host symmetry-protected Majorana fermions at the boundary, and upon tuning through a time-reversal breaking (TRB) quantum phase transition, the critical Ising-type order parameter will inevitably couple to the boundary Majorana fermion.  
The simultaneous transition of bulk and boundary occurs at a multi-critical point, where the Yukawa coupling between the $d$-dimensional Ising BCFT and the $(d-1)$-dimensional boundary Majorana will drive the combined system to a nontrivial boundary GNY fixed point. 
While one-loop renormalization group (RG) analyses have explored this fixed point~\cite{shen2024new}, its realization in microscopic lattice models and the corresponding nonperturbative investigation remain open questions.

\begin{figure}
    \includegraphics[width=0.8\linewidth]
    {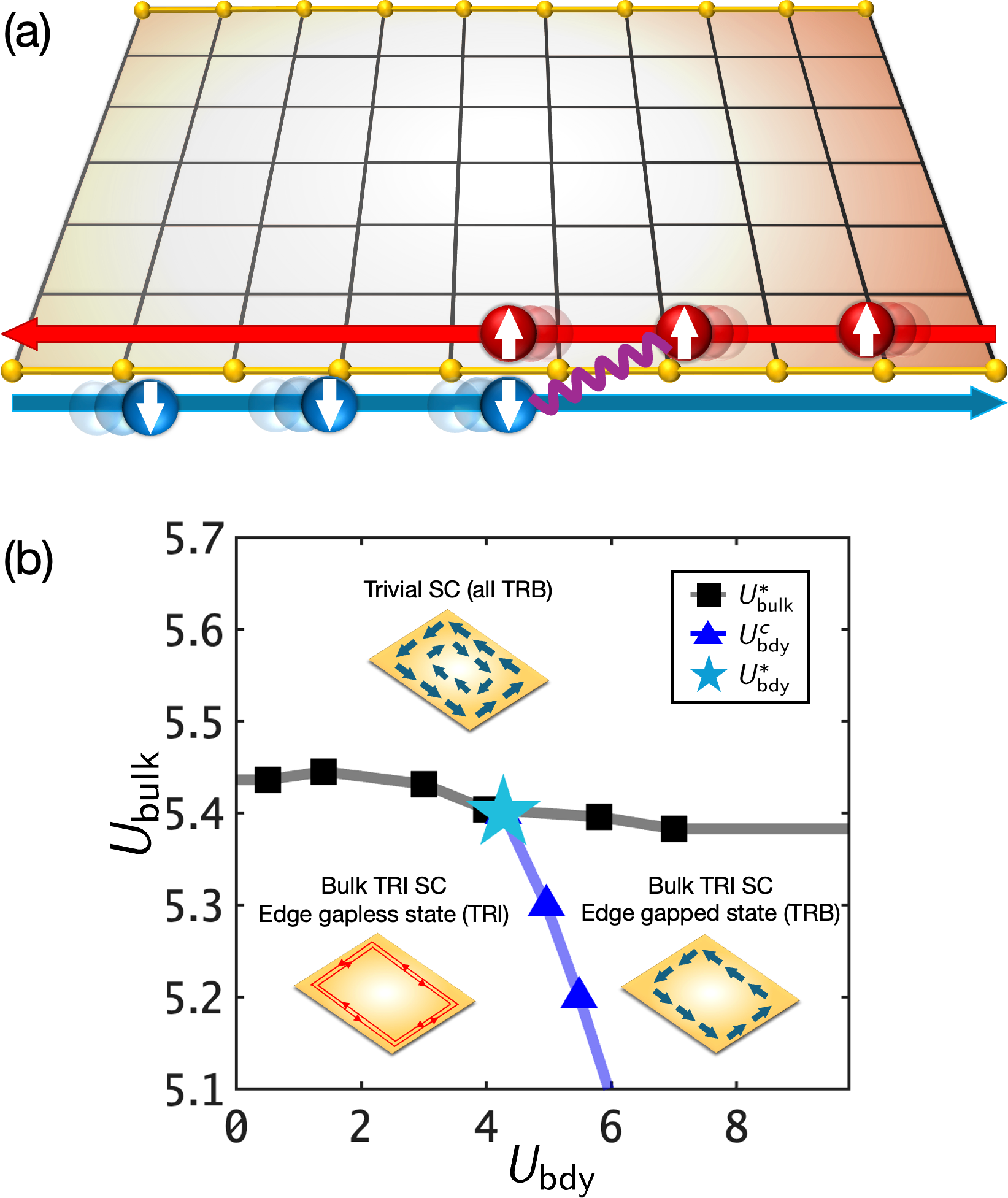}
    \caption{(a) Illustration of the square lattice with open boundaries at the top and bottom. Bulk bonds are shown in black, boundary bonds in yellow. In the disordered phase, the open boundaries host gapless helical Majorana modes subjected to an attractive Hubbard interaction.
    (b) Phase diagram of the lattice model in Eq.~\eqref{eq:QMC_lattice}. The gray line traces the bulk transition, the blue curve the surface transition, and the star marks the special point.}
    \label{fig:lattice-model}
\end{figure}

To address this question, we consider a microscopic model of spinful electrons on a two-dimensional square lattice %
featuring a nontrivial TRI topological pairing with spin-up (spin-down) electrons forming $p_x+ip_y$ ($p_x-ip_y$) pairings. 
The model falls in the DIII SPT class~\cite{ryu2010topological} protected by time-reversal symmetry, hosting helical Majorana fermions on the boundary in the topological phase, as illustrated in Fig.~\ref{fig:lattice-model}(a). 
To induce a bulk quantum phase transition, we incorporate onsite attractive Hubbard interactions that favor the TRB $s$-wave pairing into the lattice model.  
Consequently, when the Hubbard interactions are sufficiently large, the system transitions into a trivial TRB superconducting phase, in which the helical Majorana edge states become fully gapped~\cite{li2017edge2}.

At the bulk quantum critical point, we vary the Hubbard interaction on the boundary independently to reveal the boundary critical phenomena. 
In particular, using large-scale
determinant quantum Monte Carlo (DQMC) simulations~\cite{blankenbeckler1981monte,hirsch1985two,assaad2008world,assaad2020ALF}, we identify distinct boundary transitions, including ordinary, special, and extraordinary phase transitions. 
The critical behavior at the special transition is characterized by an additional fermionic critical exponent, distinguishing it from all previously known boundary universality classes and thereby identifying it as the boundary GNY universality class.

To further demonstrate consistency with theory, we compute the critical exponents via a two-loop RG calculation. 
Remarkably, the critical exponents obtained from the RG analysis are consistent with those from the DQMC simulation, firmly establishing the boundary GNY universality class in a lattice model. 
Finally, we briefly discuss the potential experimental realizations of the boundary GNY universality class and outline directions for future investigations.

\vspace{10pt}
\noindent{\bf Determinant quantum Monte Carlo simulations} \\
The full model on the square lattice consists of a non-interacting Hamiltonian $H_{\rm TSC}$ that features $p_x+ip_y$ pairing for the spin-up electrons and $p_x-ip_y$ for the spin-down electrons, and an attractive Hubbard interaction $H_{U}$ that drives a quantum phase transition into a TRB singlet pairing:
\begin{eqnarray}
    \label{eq:QMC_lattice}
    && H=H_{\text{TSC}}+H_U \,, \\
    && H_{\text{TSC}}=-t\sum_{\langle \boldsymbol i\boldsymbol j\rangle ,\sigma}\left(c_{\boldsymbol i\sigma}^\dagger c_{\boldsymbol j\sigma}+h.c.\right)-\mu\sum_{\boldsymbol i,\sigma}c_{\boldsymbol i\sigma}^\dagger c_{\boldsymbol i\sigma} \\
    && \quad -\Delta \left(\sum_{\langle \boldsymbol i\boldsymbol j\rangle_x,\sigma}c_{\boldsymbol i\sigma}^\dagger c_{\boldsymbol j\sigma}^\dagger+i\sum_{\langle \boldsymbol i\boldsymbol j\rangle_y,\sigma}(-1)^\sigma c_{\boldsymbol i\sigma}^\dagger c_{\boldsymbol j\sigma}^\dagger+\text{H.c.}\right)\,,\nonumber\\
    && H_U=-U_{\text{bulk}}\sum_{\boldsymbol i\in{\text{bulk}}}\left(c_{\boldsymbol i\uparrow}^\dagger c_{\boldsymbol i\uparrow}-\frac12\right)\left(c_{\boldsymbol i\downarrow}^\dagger c_{\boldsymbol i\downarrow}-\frac12\right)\nonumber\\
    && \quad -U_{\text{bdy}}\sum_{\boldsymbol i\in{\text{bdy}}}\left(c_{\boldsymbol i\uparrow}^\dagger c_{\boldsymbol i\uparrow}-\frac12\right)\left(c_{\boldsymbol i\downarrow}^\dagger c_{\boldsymbol i\downarrow}-\frac12\right)\,,
\end{eqnarray}
where $c_{\boldsymbol{i}\sigma}$ ($c_{\boldsymbol{i}\sigma}^\dagger$) annihilates (creates) a electron of spin $\sigma$ on site $\boldsymbol{i}$, while $t$ and $\mu$ are the hopping amplitude and the chemical potential, respectively. 
The pairing amplitude $\Delta$ is real. The pairing phase between nearest-neighbor sites $\langle \boldsymbol i\boldsymbol j\rangle$ changes by $\pi/2$ from the bonds along the $x$ direction to those along the $y$ direction, denoted by $\langle \boldsymbol i\boldsymbol j\rangle_x$ and $\langle \boldsymbol i\boldsymbol j\rangle_y$, respectively. 
The phase also differs in sign $(-1)^{\sigma}\equiv \pm 1$ for spin-up and spin-down sectors, respectively. 
The parameters $U_{\text{bulk}}$ and $U_{\text{bdy}}$ are the on-site attractive Hubbard interaction strengths in the bulk and on the boundary, respectively. The bulk (boundary) lattice is depicted in Fig.~\ref{fig:lattice-model}(a) in black (yellow). 
By tuning $U_{\text{bulk}}$ and $U_{\text{bdy}}$, we can map out the boundary quantum phase diagram.

The topological Majorana edge state in the TSC phase is protected by the time-reversal symmetry~\cite{ryu2010topological}, denoted by $\mathcal T=i\sigma^y\mathcal K$, where $\sigma^y$ is the Pauli matrix acting on spin, and $\mathcal K$ is the complex conjugate operator. The Hamiltonian also preserves another time reversal symmetry $\mathcal{\tilde T}=\sigma^x \mathcal{K}$.   
The presence of both $\mathcal T$ with $\mathcal T^2=-1$ and $\mathcal{\tilde T}$ with $\tilde{\mathcal T}^2=1$ allows a sign-problem-free formulation of DQMC in the Majorana class~\cite{li2015majorana,li2016majorana,wei2016majorana,li2019annualrev}. 
Increasing the onsite Hubbard interaction promotes the trivial $s$-wave pairing, $\Delta_{\boldsymbol{i}}^s\equiv ic_{\boldsymbol{i}\uparrow}^\dagger c_{\boldsymbol{i}\downarrow}^\dagger+\text{H.c.}$, that spontaneously breaks the time-reversal symmetry, as $\mathcal T\Delta^s\mathcal T^{-1}=-\Delta^s$. 

\begin{figure}
    \includegraphics[width=0.99\linewidth]{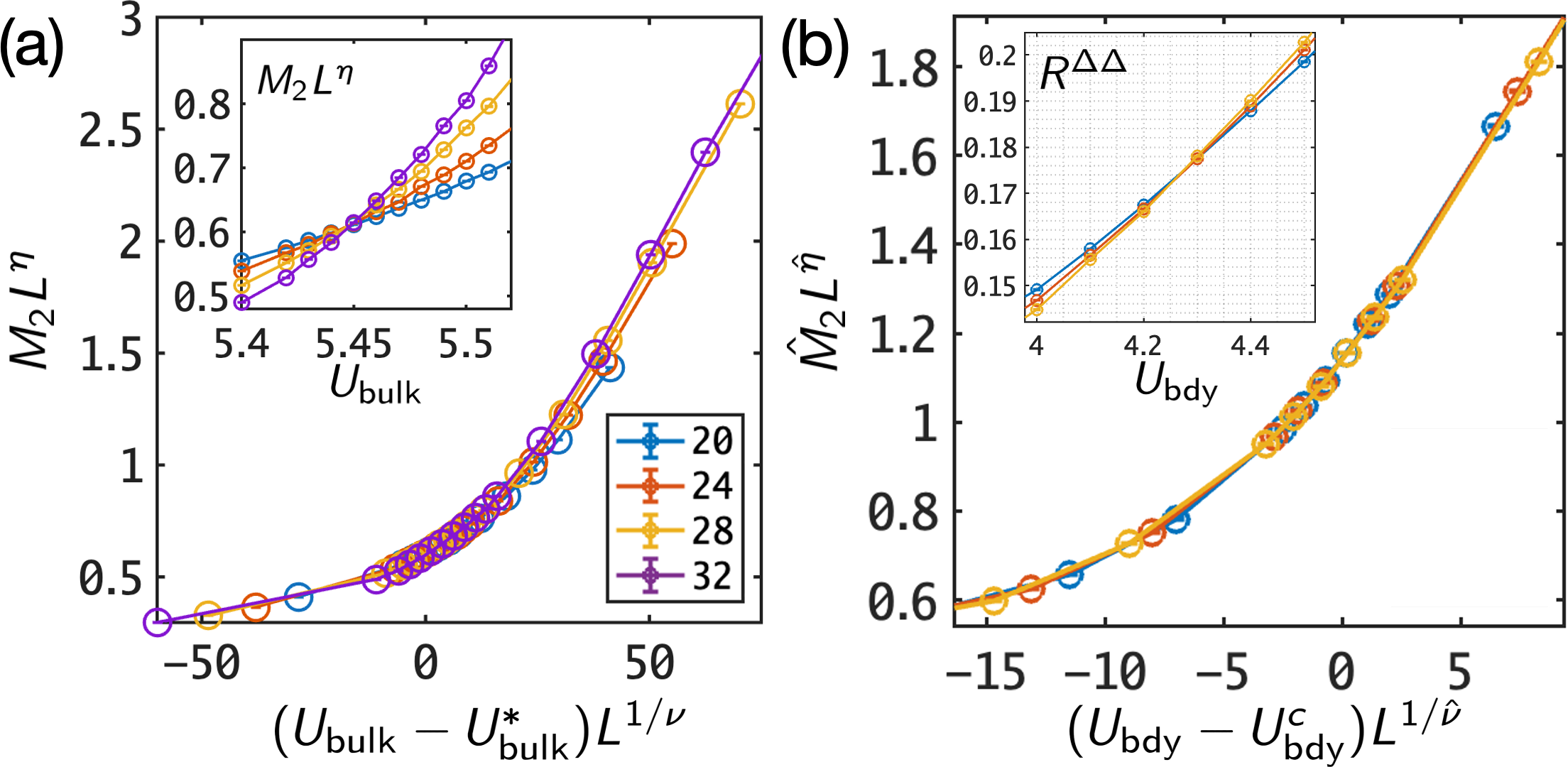}
    \caption{(a) Scaling collapse of the bulk $M_2$ as a function of $U_\text{bulk}$ near the bulk criticality at $U_\text{bdy}=1.4$. Here $\eta=1.037$, and $\nu=0.630$. Inset shows the crossing of $M_2L^\eta$ for different system sizes at the bulk transition point $U^*_\text{bulk}$. 
    (b) Scaling collapse of the boundary $\hat{M}_2$ as a function of $U_\text{bdy}$ near the boundary criticality at $U_\text{bulk}=5.4$. Here $\hat\eta=0.693$, and $\hat{\nu}=1.37$. Inset shows the crossing of the RG invariant at $U^c_\text{bdy}=U^*_\text{bdy}=4.28$.}\label{fig:collapse-qcp}
\end{figure}

To explore the quantum phase diagram, we employ the nonperturbative large-scale DQMC. 
We use a cylinder lattice geometry that is periodic in $x$, with the number of unit cells along the $x$ and $y$ being $L\equiv L_x=2L_y$. 
The computed phase diagram is displayed in Fig.~\ref{fig:lattice-model}(b), exhibiting boundary criticality with ordinary, special, and extraordinary phase transitions. 
The parameters in $H_\text{TSC}$ are set to $t=1$, $\mu=-0.5$, and $\Delta=0.4$. 
The simulated sizes range from $L=20$ to $L=32$, with a large inverse temperature from $\beta=85$ to $\beta=120$ to obtain the phase diagram at zero temperature. 
The bulk phase transition point $U_{\rm bulk}^*$ is determined by the crossing of the order parameter moment, defined as
\begin{eqnarray}
    \label{eq:RG_m2}
    M_2\equiv \frac1{L_x^2 L_y^2}\sum_{\boldsymbol{i},\boldsymbol{j}}\langle\Delta_{\boldsymbol{i}}^s\Delta_{\boldsymbol{j}}^s\rangle\,.
\end{eqnarray} 
In the scaling limit, the bulk $M_2$ behavior approaches $M_2 \sim L^{-\eta }f\left[\left(U-U^*\right)L^{1/\nu}\right]$, where $\nu$ and $\eta$ are the critical exponents. 
It is well known that the bulk critical point belongs to the 3D Ising universality class. 
Thus, using the $\eta\approx 1.037$ from the 3D Ising universality class, the bulk $M_2$ crosses at  $U_{\text{bulk}}^* \approx 5.45$, as shown in the inset of Fig.~\ref{fig:collapse-qcp}(a), signaling a bulk quantum phase transition. 
Further finite-size scaling (FSS) analysis using $\nu \approx 0.630 $ leads to the excellent data collapse in the main panel of Fig.~\ref{fig:collapse-qcp}(a). 
Note that for larger $U_{\text{bdy}}$ the crossing of $M_2$ is affected by the surface transition in finite-size systems and shifted downwards, as seen in Fig.~\ref{fig:lattice-model}(b).

Next, we investigate the boundary phase transition. 
The boundary phase transition occurs before the bulk becomes ordered, $U_{\text{bulk}}<U_{\text{bulk}}^*$, when $U_\text{bdy}$ is sufficiently large. 
The boundary critical point, $U_{\rm bdy}^c$, is located at the crossing of the dimensionless RG-invariant quantity given by the spatial correlator of the order parameter at one edge of the cylinder. It is defined by
\begin{eqnarray}
    \label{eq:RG_ratio}
    R_{\text{bdy}}^{\Delta\Delta}\equiv\frac{1}{2\pi}\sqrt{\left|\tilde C_{\rm bdy}^{\Delta\Delta}(0)/{\text{Re}}\tilde C_{\text{bdy}}^{\Delta\Delta}(k_{\text{min}})\right|-1} \,,
\end{eqnarray}
where the minimal momentum is $k_{\text{min}}=2\pi/L_y$, and the correlator is
\begin{align}
    \tilde{C}_{\text{bdy}}^{\Delta\Delta}(k)&=\sum_{\boldsymbol{i},\boldsymbol{j}\in\text{bdy}}\langle\Delta_{\boldsymbol{i}}^s\Delta_{\boldsymbol{j}}^s\rangle e^{-ik(x_{\boldsymbol i}-x_{\boldsymbol j})}.
\end{align}
where, $\boldsymbol{i}=(x_{\boldsymbol{i}},y_{\boldsymbol{i}})$, and $\boldsymbol{j}=(x_{\boldsymbol{j}},y_{\boldsymbol{j}})$ are the
displacements to lattice sites. 
$U^{c}_\text{bdy}$ extracted from the RG-invariant quantity for $U_{\text{bulk}} < U_{\text{bulk}}^\ast$ defines the surface transition, as shown by the blue curve in Fig.~\ref{fig:lattice-model}(b). 
Interestingly, the surface transition features an emergent (1+1)-dimensional supersymmetry~\cite{friedan1984superconformal,grover2013emergent,fei2016yukawa,li2017edge2}.

As $U_\text{bulk}$ increases, $U^c_\text{bdy}$ decreases. 
Eventually, the surface transition line
merges with the bulk transition point, marking the special transition, as indicated by the star in Fig.~\ref{fig:lattice-model}(b). 
At the bulk phase transition line $U_{\text{bulk}}=U_{\text{bulk}}^*$, the boundary Majorana fermion develops a finite mass in the thermodynamic limit when the boundary interaction strength $U_{\text{bdy}}$ exceeds a critical value $U_{\text{bdy}}^*$. %
Hence, for $U_{\text{bdy}}<U_{\text{bdy}}^*$ ($U_{\text{bdy}} > U_{\text{bdy}}^\ast$) the boundary Majorana fermion is gapless (gapped), corresponding to an ordinary (extraordinary) phase transition. 
The intersection between the surface transition, the ordinary transition, and the extraordinary transition, indicated by the star at $(U^*_\text{bdy},U^*_\text{bulk}) \approx (5.4, 4.28)$ in Fig.~\ref{fig:lattice-model}(b), is precisely the special transition.   

Having revealed the boundary quantum phase diagram, we are ready to extract quantitatively the critical exponent at the special transition. 
Using the surface $\hat M_2$, defined similarly to Eq.~\eqref{eq:RG_m2}, but summing over all sites on a single edge only, i.e., 
\begin{eqnarray}
    \hat M_2 \equiv \frac1{L_x^2}\sum_{\boldsymbol{i},\boldsymbol{j} \in {\rm bdy}}\langle\Delta_{\boldsymbol{i}}^s\Delta_{\boldsymbol{j}}^s\rangle\,,
\end{eqnarray}
we perform the FSS to obtain the boundary critical exponents $\hat\eta$ and $\hat\nu$. 
The scaling collapse of $\hat{M}_2$ in Fig.~\ref{fig:collapse-qcp}(b) gives the boundary critical exponents,
$\hat\eta=0.693(6)$ and $\hat\nu=1.37(5)$, at the special transition. 
These are consistent with the 3D Ising BCFT ($\hat\eta_\text{ref}=0.728$ and $\hat\nu_\text{ref}=1.400$ reported in Monte Carlo simulation~\cite{deng2005surface}). %

\begin{figure}[t]
	\includegraphics[width=0.99\linewidth]{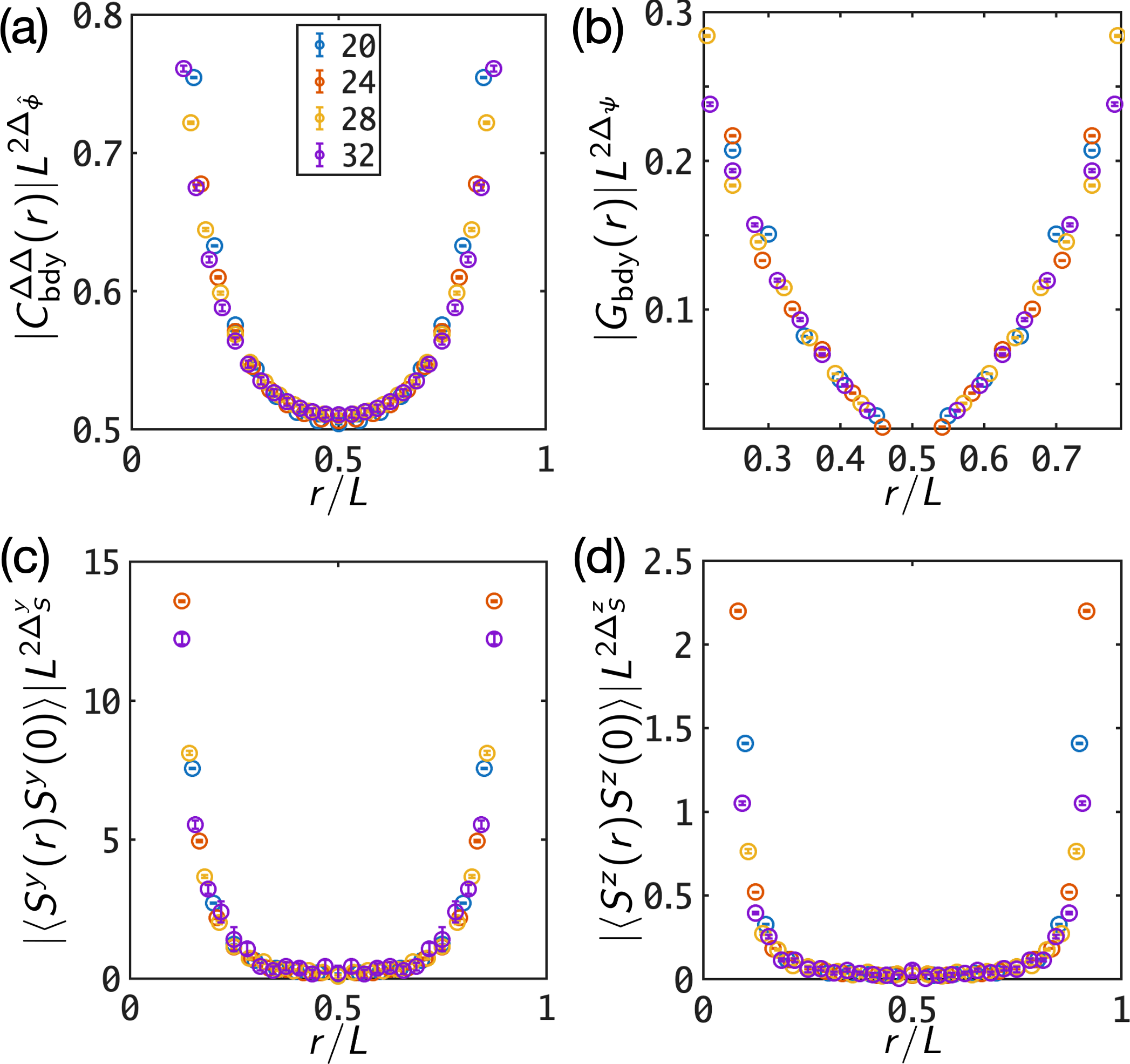}
	\caption{(a)--(b) Scaling collapse of the boundary (a) boson correlator and (b) fermion correlator, at $U^*_\text{bulk}$ and $U^*_\text{bdy}$. Here, $\Delta_{\hat\phi}=0.33$ and $\Delta_\psi=0.58$. (c)--(d) Scaling collapse of the boundary spin correlators (c) $S^y$ and (d) $S^z$. Here, $\Delta^y_S=1.62$ and $\Delta^z_S=1.26$.}
	\label{fig:special}
\end{figure}

Therefore, at the special transition point, the boundary order parameter has the same scaling dimension as in the Ising BCFT.
This is further confirmed by the FSS analysis in the following numerics and the RG calculation in the next section. 
At the transition point, the correlation function of the boundary order parameter exhibits a scaling form,
\begin{eqnarray} \label{eq:cft_ansatz}
C_{\text{bdy}}^{\Delta\Delta}(r)  \sim  L^{-2\Delta_{\hat\phi}}f\left[\sin\left(\pi r/L\right)\right] \,,
\end{eqnarray}
Extracted from the scaling collapse in Fig.~\ref{fig:special}(a), the scaling dimension of the boundary boson is $\Delta_{\hat\phi}=0.33(2)\approx\hat\eta/2$, consistent with the value of $\Delta_\text{sp}\approx0.364$ obtained in the boundary 3D Ising model~\cite{deng2005surface}. 
However, the presence of the boundary Majorana state leads to an additional fermion boundary critical exponent. 
With a similar scaling ansatz for the fermion correlation function, $G_{\rm bdy}(r) \sim L^{-2\Delta_\psi} f\left[ \sin(\pi r)/L \right]$, we show the data collapse yielding $\Delta_{\psi}=0.58(5)$ at the special transition.
This additional fermionic boundary critical exponent, distinct from all previously known boundary fixed points, provides definitive evidence for the boundary GNY universality class! 
For completeness, we note that the correlation function of the boundary order parameter and the boundary fermion are defined by $C_{\text{bdy}}^{\Delta\Delta}(r) \equiv \sum_{\boldsymbol{l}\in \text{bdy}} \langle \Delta_{\boldsymbol{l}}^s \Delta^s_{\boldsymbol{l} + \boldsymbol{r}} \rangle/L $ and $G_{\text{bdy}}(r)\equiv\sum_{\boldsymbol{l}\in \text{bdy}}\langle c_{\boldsymbol{l}}^\dagger c_{\boldsymbol{l}+\boldsymbol{r}}\rangle/L$, respectively. 
Here, $\boldsymbol{r} = (r,0)$.

\vspace{10pt}
\noindent{\bf Renormalization group analysis} \\
At the bulk quantum critical point, the system is described by a $d$-dimensional Ising order parameter $\phi$, coupled to a $(d-1)$-dimensional boundary Majorana fermion $\psi$ through a Yukawa interaction. 
The low-energy effective action consists of the bosonic part, the fermionic part, and the interaction part: $S=S_{\rm b}+S_{\rm f}+S_{\rm int}$.

The bosonic action is the $d$-dimensional Ising BCFT, 
\begin{align}
    S_{\rm b}=\int_{\mathcal M}{\rm d}^dx\left[\frac{1}{2}\left(\partial_\mu\phi\right)^2+\frac{\lambda}{4!}\phi^4\right]+\int_{\partial\mathcal M}{\rm d}^{d-1}x\,h\phi^2,
\end{align}
where $\partial_\mu$ denotes derivatives with respect to (imaginary) time $x_0$ and space $x_i$ components, $i=1,2,\dots, d-1$. 
The field theory lives in a $d$-dimensional semi-infinite spacetime, $\mathcal M=\{x_\mu \mid x_2>0\}$, with the boundary at $\partial\mathcal M=\{x_\mu \mid x_2=0\}$. 
The scalar field $\phi$ is odd under the time-reversal symmetry. 
The parameter $\lambda$ denotes the quartic boson self-interaction in the bulk, while the bilinear boundary term $h\phi^2$ controls the surface universality class: ordinary ($h>0$), special ($h=0$), or extraordinary ($h<0$).

The $(d-1)$-dimensional boundary theory consists of $N$ flavors of four-component Dirac fermions $\psi_i$ (equivalently, $4N$ flavors of two-component Majorana fermions), $i=1,2,\dots, N$:
\begin{eqnarray} 
\label{eq:Sf}
    S_{\text{f}} = \int_{\partial\mathcal M}{\rm d}^{d-1}x\,\bar\psi_i\gamma^\mu\partial_\mu\psi_i \,,
\end{eqnarray}
where $\gamma^\mu$ denotes the Gamma matrix $\{\gamma^\mu, \gamma^\nu \} = 2 \delta^{\mu\nu}$, $\bar\psi_i=\psi_i^\dagger\gamma^0$. 
The kinetic term is $\gamma^\mu \partial_\mu \equiv \gamma^0 \partial_0 + \gamma^1 \partial_1 + \gamma^3 \partial_3 + \dots + \gamma^{d-1} \partial_{d-1}$, while $x_2 = 0$ for the boundary. 
For the explicit calculation, we use four-dimensional Gamma matrices, whereby $\Tr[\gamma^\mu \gamma^\nu] = 4 \delta^{\mu\nu}$. 
Note that the boundary Majorana fermion in the 2D TSC simulated in our DQMC is a single-flavor two-component fermion, corresponding to $N=1/4$.

The Ising BCFT is coupled to the boundary Majorana field theory via a symmetry-allowed Yukawa coupling: 
\begin{eqnarray}
\label{eq:yukawa}
     S_{\text{int}} =\int_{\partial\mathcal M}{\rm d}^{d-1}x\, g\phi\bar\psi_i\psi_i \,.
\end{eqnarray} 
The Yukawa coupling can generate a new universality class.
Hence, we treat it as a perturbation and first analyze its fate. 
Since the DQMC simulation is implemented on the $(2+1)$-dimensional TSC model, we focus on $d=3$.
The Majorana fermion in Eq.~\eqref{eq:Sf} has a scaling dimension $(d-2)/2 = 1/2$. 
Then, according to Eq.~\eqref{eq:yukawa}, the Yukawa coupling is irrelevant (relevant) if the scaling dimension of the boundary order parameter is greater (less) than one. 
The scaling dimension of the boundary order parameter depends on the boundary universality class~\cite{deng2005surface}. 
For the ordinary transition, its scaling dimension is approximately $\Delta_\text{ord} \approx 1.263$.  
This renders the Yukawa coupling irrelevant. 
Hence, the boundary critical theory reduces to the decoupled free Majorana theory and Ising BCFT. 
For the extraordinary transition, the boundary order parameter acquires a finite expectation value spontaneously, gapping out the boundary Majorana fermion. 
The boundary critical theory solely consists of the Ising BCFT in the low-energy limit. 
Most significantly, at the special transition, the boundary order parameter has a scaling dimension less than one, i.e., $\Delta_\text{sp} \approx 0.364$, making the Yukawa coupling a relevant perturbation.

Therefore, we focus on the special transition point at $h=0$.
To perform a controlled calculation, we employ the $d= 4 -\epsilon$ expansion with in the RG framework. 
It is easy to check that both coupling strengths $\lambda$ and $g$ are marginal in $d=4$ at the tree level. %
The pioneering work of Ref.~\onlinecite{shen2024new} carried out a one-loop analysis of this theory and identified a new fixed point, dubbed the boundary GNY universality class. 
To improve the estimate of critical exponents, we perform a two-loop expansion in the minimal subtraction scheme. 
After extensive calculation, the RG equation of the Yukawa coupling reads
\begin{eqnarray}
    \label{eq:eg_g}
        \frac{{\rm d}g}{{\rm d}\log\mu}=-\frac\epsilon2 g +f_0g^3+f_1g\lambda+f_2g^5 +f_3g^3\lambda+f_4g\lambda^2 \,, \nonumber \\
\end{eqnarray}
where the coefficients $f_i$ are given in the Supplementary Information, together with a detailed derivation.
The anomalous dimension for the Majorana fermion is also computed to the two-loop order,
\begin{eqnarray}
    \eta_{\psi} &=& \frac{g^2}{12\pi^2}+\frac{g^2\lambda\left(32-9\gamma_\text{E}+9\log\pi\right)}{1152\pi^4}\\
    && -\frac{g^4\left(6N\pi^2+42+16\gamma_\text{E}-140\log2-16\log\pi\right)}{288\pi^4}\, ,\nonumber
\end{eqnarray}
where $\gamma_\text{E}$ is the Euler-Mascheroni constant. 

\begin{table}[t]
\centering
\begin{tabular}{lll}
\hline\hline
Method            & $\Delta_{\hat\phi}$      &   $\Delta_\psi$   \\[2pt]
\hline 
$4-\epsilon$ (two-loop)   & 0.340     & 0.622 \\[2pt]
DQMC         & 0.33(2)  & 0.58(5) \\
[2pt]
\hline\hline
\end{tabular}
\caption{The critical exponents for the boundary order parameter $\Delta_{\hat \phi}$ and the boundary Majorana fermion $\Delta_{\psi}$ for the $d=3$ boundary GNY universality class, from the $d = 4-\varepsilon$ RG calculation and the DQMC simulation, respectively.}
\label{tab:critical_exponents}
\end{table}

Because the boundary Majorana fermion will not affect the bulk Wilson-Fisher (WF) fixed point, we can fix the quartic bulk coupling strength at the WF fixed point, $\lambda^*=\frac{16\pi^2}{3}\epsilon+\frac{2448\pi^2}{1331}\epsilon^2$. 
With this, the RG equation for the Yukawa coupling admits a nontrivial fixed point with $g^{\ast 2} > 0$, corresponding to the boundary GNY universality class. 
By setting $\epsilon=4-d=1$, and $N=1/4$ for a single two-component boundary Majorana fermion, we find that, at the boundary GNY fixed point, the two-loop scaling dimension for the boundary Majorana fermion is $\Delta_{\psi} = 1/2 + \eta_\psi \approx 0.622$, and for the boundary order parameter is $\Delta_{\hat\phi}\approx0.340$. 
These values are consistent with the DQMC simulation results, as summarized in Table~\ref{tab:critical_exponents}. 
Notice that the scaling dimension of the boundary order parameter remains unchanged at two-loop order, even in the presence of a finite Yukawa coupling. 
This is evident from the dimensional analysis of loop diagrams involving the fermion propagator: all the integrals are convergent.

\vspace{10pt}
\noindent{\bf Discussion} \\
We have investigated the boundary criticality of two-dimensional correlated topological superconductors through large-scale DQMC and two-loop RG analysis, firmly establishing the boundary GNY universality class. 
To our knowledge, this represents the first nonperturbative study of boundary criticality in two-dimensional TRI topological superconductors. 
The comparison between the DQMC simulations and the RG analysis shows that the two methods yield consistent quantitative results for the boundary critical exponents. 
Taken together, these findings provide strong evidence that the special transition of the TRI topological superconductor falls into the boundary GNY universality class.

In addition to the bosonic and fermionic critical exponents, the boundary spins, $S^\mu_{\boldsymbol{i}}=\frac12 c^\dagger_{\boldsymbol i}\sigma^\mu c_{\boldsymbol{i}}$, also offer a direct probe of the magnetic response at the edge, where $\sigma^\mu$ are the Pauli matrices in the spin space. 
We determine its scaling nonperturbatively via DQMC simulations of boundary spin correlations. 
As shown in Fig.~\ref{fig:special}(c) and (d), the exponents are $\Delta_{S}^y=1.62(7)$ and $\Delta_{S}^z=1.26(9)$, corresponding to the spin components along the $y$ and $z$ directions, respectively. 
Complementarily, we perform a perturbative RG for the spin operator (see the Supplementary Information).
The spin operator $S^y$ and $S^z$ correspond to the $i\bar\psi\gamma^5\psi$ and $i\bar\psi\gamma^1\psi$ vertices, respectively, in the effective field theory. 
Note that the $S^x$ lies in the same irreducible representation of the TRB superconducting order, so it exhibits the same critical exponent as $\hat \phi$. 
At the one-loop level, the RG calculation yields $\Delta_{S}^y=d-2+\frac{2\epsilon}{3}\approx1.67$, and $\Delta_{S}^z=d-2+\frac{\epsilon}{3}\approx1.33$. 
Our RG analysis framework can be systematically extended to higher orders, and we provide the details in the Supplementary Information. 
Notably, the DQMC and RG calculations are fully consistent and also agree on the anisotropy of the spin channels, thus offering a robust theoretical picture that can guide the experiment on boundary spins in a TRI topological superconductor.

The interaction effect of TRI topological superconductors can lead to a novel topological classification~\cite{qi2013new,ryu2012interacting,yao2013interaction}. 
It would be interesting to investigate the boundary critical phenomena for different $N$ (note that in our convention, $4N$ is the number of two-component Majorana fermions). 
As the $N$ dependence of the fermion boundary critical exponent is manifested in our two-loop calculation (see Supplementary Information), it is desirable to generalize the DQMC simulation for different $N$ to extract the exponents nonperturbatively.  
Finally, we point out that iron-chalcogenide superconductors, such as FeSe$_{1-x}$Te$_x$, provide a promising setting for exploring the boundary critical phenomena~\cite{shibauchi2020exotic}. 
Recent $\mu$SR experiments~\cite{roppongi2025topology} have reported bulk time-reversal symmetry breaking in compositions that also host topological surface states~\cite{zhang2018observation,wang2018evidence}. 
This coexistence suggests that FeSe$_{1-x}$Te$_x$ could serve as a candidate material for the experimental realization of the boundary GNY universality class. 

\vspace{10pt}
\noindent \textbf{Acknowledgments:}
We thank Zi-Xiang Li and Zhou-Quan
Wan for helpful discussions. DQMC simulations used the Algorithms for Lattice Fermions (ALF) package \cite{assaad2020ALF}. 
Scaling analysis is aided by autoScale.py \cite{autoscale}. 
The numerical calculation was performed using high-performance computational resources (HPC) provided by the Louisiana Optical Network Infrastructure. 
This work is supported in part by a start-up fund (Y.G., H.J., and S.-K.J.), and the Lavin-Bernick grant (Y.G.) from Tulane University, MOSTC under Grant No.~2021YFA1400100 (H.Y.), NSFC under Grant Nos.~12347107 and 12334003
(H.Y.), and the New Cornerstone Science Foundation through the Xplorer Prize (H.Y.).

\bibliography{reference}

\clearpage
\onecolumngrid   %

\begin{center}
\textbf{\large Supplementary Information}
\end{center}

\setcounter{secnumdepth}{3}

\setcounter{section}{0}

\setcounter{equation}{0}
\renewcommand\theequation{S\arabic{equation}}

\setcounter{figure}{0}
\renewcommand\thefigure{S\arabic{figure}}
\renewcommand{\figurename}{Supplementary Figure}

\setcounter{table}{0}
\renewcommand\thetable{S\arabic{table}}
\renewcommand{\tablename}{Supplementary Table}

\section{Determinant quantum Monte Carlo simulation}
\begin{figure}[h]
\includegraphics[width=0.28\textwidth]{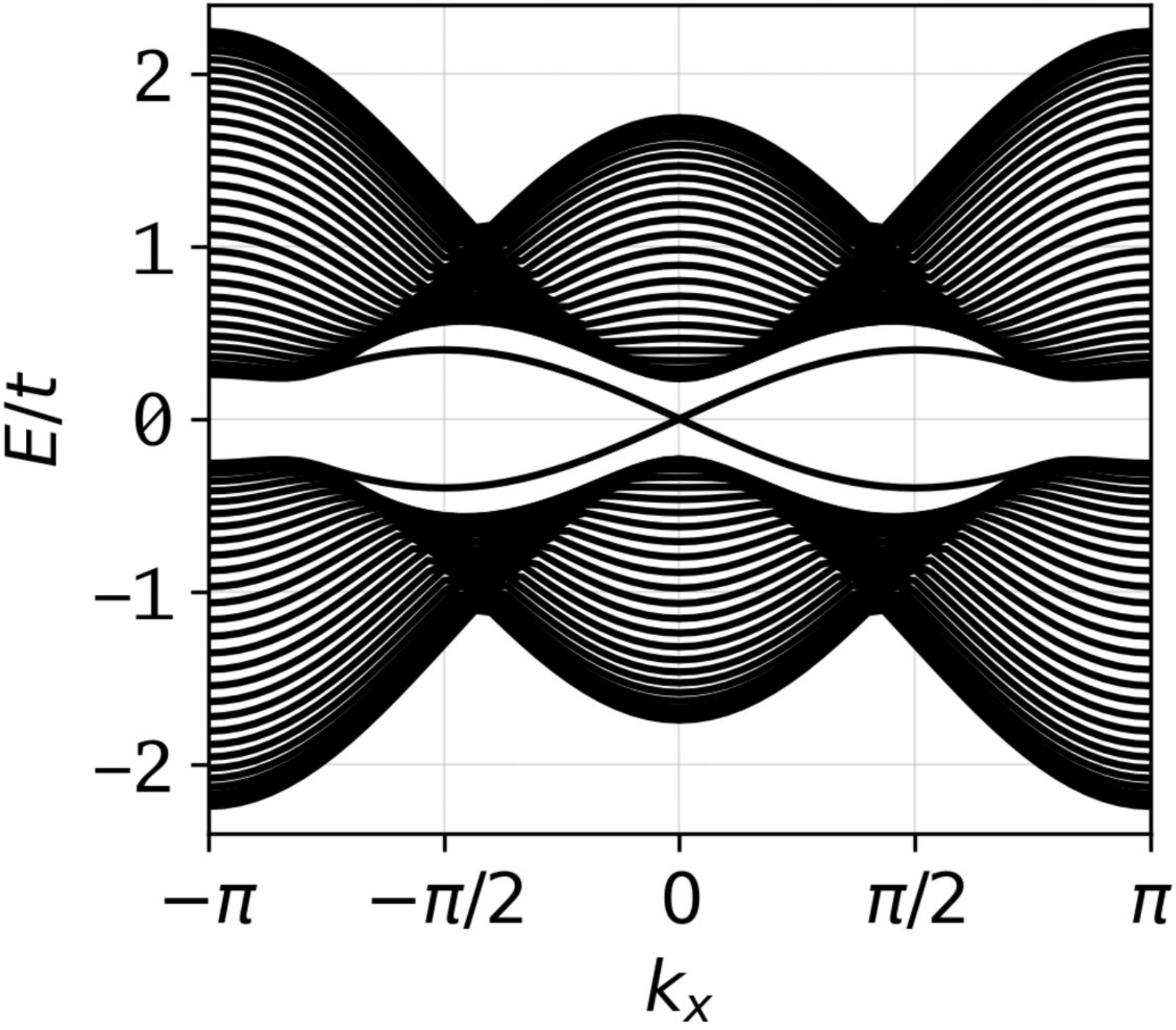}
\caption{\label{fig:h0} The energy dispersion of the noninteracting $H_\text{TSC}$ in the topological regime, with parameters used in the paper. Gapless edge Majorana modes run inside the bulk gap.}
\end{figure}
We perform finite-temperature determinant quantum Monte Carlo (DQMC) simulations using the ALF package~\cite{assaad2020ALF}.
The Bogoliubov de-Gennes Hamiltonian $H_\text{TSC}$ is represented in the Nambu-spinor basis $(c_{\boldsymbol i\sigma}\ d_{\boldsymbol i\sigma})^T$ where $d_{\boldsymbol i\sigma}\equiv c^\dagger_{\boldsymbol i\sigma}$. 
The spectrum of $H_\text{TSC}$ is plotted in Fig.~\ref{fig:h0} in the Hilbert space of Nambu spinors.

The onsite Hubbard interaction is decoupled via the discrete Hubbard-Stratonovich transformation, which approximates the auxiliary field integral with the four-point Gauss-Hermite quadrature, by summing over the discrete auxiliary fields $x_j$ at the roots of the Hermite polynomial labeled by $j=\pm1,\pm2$, together with their weights $w_j$,
\begin{eqnarray}
&&e^{U_{\boldsymbol i}(n_{\boldsymbol i\uparrow}-\frac12)(n_{\boldsymbol i \downarrow}-\frac12)\mathrm{d}\tau-\frac14 U_{\boldsymbol i} \mathrm{d}\tau}
\approx 
\frac14\sum_{j=\pm1,\pm2} w_j e^{x_j\frac{\sqrt{-U_{\boldsymbol i}\mathrm{d}\tau}}{2}\sum_\sigma (-1)^\sigma(c^\dagger_{\boldsymbol i\sigma}c_{\boldsymbol i\sigma}-d^\dagger_{\boldsymbol i\sigma}d_{\boldsymbol i\sigma})}\,,
\end{eqnarray}
where $(-1)^\sigma=\pm$ for the spin-up terms and the spin-down terms, respectively.
The weights are $w_{\pm1}=1+\sqrt{6}/3$ and $w_{\pm2}=1-\sqrt{6}/3$, at the roots $x_{\pm1}=\pm\sqrt{2\left(3-\sqrt6\right)}$ and $x_{\pm2}=\pm\sqrt{2\left(3+\sqrt6\right)}$~\cite{assaad2020ALF}.

The decoupling scheme above preserves the time-reversal symmetry $\mathcal{T}$, which is a bijection between the two spin sectors. As discussed in the main text, the presence of two Majorana time-reversal symmetries, with $\mathcal{T}^2=-1$ for one, ensures that the Hamiltonian is sign-problem-free~\cite{li2016majorana,li2017edge2}.
Due to the bijection, the DQMC sampling is only needed in one spin sector, while the sampling in the other sector follows from time-reversal symmetry. 
The absolute value of the determinant weight of one spin sector in the Nambu-spinor basis is exactly that of the entire system, due to the doubling of the Hilbert space. 
The Green's functions of spin-down fermions are simply the complex conjugate of the spin-up counterparts. 
Thus, observables across both spin sectors can be computed. 
By the Wick's theorem, they are reducible to products of quadratic fermion correlators, where each correlator belongs to a single spin sector in the Monte Carlo sampling.

We use sufficiently low temperatures that the system is effectively at zero temperature for the computed observables. 
The inverse temperature is $\beta = 120/t$ for all system sizes, except that $\beta = 85/t$ for $L=20$. 
The imaginary time step is $\mathrm{d}\tau=0.05/t$.
Each DQMC simulation utilizes 8 Markov chains. 
About 400 back-and-forth sweeps are conducted in each chain for $L=32$. 
Smaller systems are simulated using more than 2000 sweeps.

Two methods of scaling collapse are used for moments of the order parameter and the spatial correlations. 
For moments of the order parameter that exhibit an inflection at the phase transition as a function of the tuning parameter, e.g., $M_2$ as a function of $U$, we perform scaling collapse with the program autoScale.py~\cite{autoscale}. 
It applies a simplex algorithm to optimize the parameters $a$, $b$, and $x_c$ in the scaling form $y = L^{-b}f[(x-x_c)L^a]$, minimizing the distance between data points $(x,y)$ for each system size $L$ and the line segments interpolating neighboring points from each of the other system sizes. This procedure also generates the uncertainty on the fitting parameters~\cite{autoscale}. 
On the other hand, for the spatial correlators, such as $G_\text{bdy}(r)\sim L^{-2\Delta_{\hat\phi}}f\left[\sin\left(\pi r/L\right)\right]$, we fit a fourth-order polynomial to the scaled variables $GL^b$ and $\sin(\pi r/L)$. 
Minimizing $\chi^2$ of the fit gives the exponent and its uncertainty. 
The polynomial fit only utilizes data at $\sin(\pi r/L)>0.5$ for bosons and $1>\sin(\pi r/L)>0.7$ for fermions to focus on the infrared behavior. The deviations of the exponents extracted from the collapses of different subsets of system sizes in a jackknife fashion produce the uncertainties.

Finally, we list the observables recorded for each in terms of single-particle Green's functions in the Nambu basis. The $s$-wave pairing is
\begin{equation}
\langle \Delta^s_{\boldsymbol{i}} \Delta^s_{\boldsymbol{j}} \rangle = | \langle
   c_{\boldsymbol{i}\uparrow}^{\dag} c_{\boldsymbol{j}\uparrow} \rangle |^2 + | \langle d_{\boldsymbol{i}\uparrow}^{\dag} c_{\boldsymbol{j}\uparrow} \rangle |^2 + |
   \langle c_{\boldsymbol{i}\uparrow}^{\dag} d_{\boldsymbol{j}\uparrow} \rangle |^2 + | \langle d_{\boldsymbol{i}\uparrow}^{\dag} d_{\boldsymbol{j}\uparrow} \rangle |^2\,,
\end{equation}
and for the spin correlators, we have
\begin{eqnarray}
  {\langle S_{\boldsymbol{i}}^x}  S_{\boldsymbol{j}}^x \rangle & = & \frac{1}{2} (\langle c_{\boldsymbol{i}\uparrow}^{\dagger} c_{\boldsymbol{j}\uparrow}
  \rangle \langle d_{\boldsymbol{i}\uparrow}^{\dagger} d_{\boldsymbol{j}\uparrow} \rangle^{\ast} - \langle d_{\boldsymbol{i}\uparrow}^{\dagger} c_{\boldsymbol{j}\uparrow}
  \rangle \langle c_{\boldsymbol{i}\uparrow}^{\dagger} d_{\boldsymbol{j}\uparrow} \rangle^{\ast}),\\
  {\langle S_{\boldsymbol{i}}^y}  S_{\boldsymbol{j}}^y \rangle & = & \frac{1}{2} (\langle c_{\boldsymbol{i}\uparrow}^{\dagger} c_{\boldsymbol{j}\uparrow}
  \rangle \langle d_{\boldsymbol{i}\uparrow}^{\dagger} d_{\boldsymbol{j}\uparrow} \rangle^{\ast} + \langle d_{\boldsymbol{i}\uparrow}^{\dagger} c_{\boldsymbol{j}\uparrow}
  \rangle \langle c_{\boldsymbol{i}\uparrow}^{\dagger} d_{\boldsymbol{j}\uparrow} \rangle^{\ast}),\\
  {\langle S_{\boldsymbol{i}}^z}  S_{\boldsymbol{j}}^z \rangle & = & \frac{1}{2} (\langle c_{\boldsymbol{i}\uparrow}^{\dagger} c_{\boldsymbol{j}\uparrow}
  \rangle \langle d_{\boldsymbol{i}\uparrow}^{\dagger} d_{\boldsymbol{j}\uparrow} \rangle + \langle c_{\boldsymbol{i}\uparrow}^{\dagger} c_{\boldsymbol{j}\uparrow}
  \rangle^{\ast} \langle d_{\boldsymbol{i}\uparrow}^{\dagger} d_{\boldsymbol{j}\uparrow} \rangle^{\ast})\,.
\end{eqnarray}
We should understand the expectation value in terms of a fixed sample, where the fermion is effectively quadratic.

\section{Renormalization group analysis}

\begin{figure}[t]
  \centering
  \subfigure[]{\includegraphics[width=0.22\textwidth]{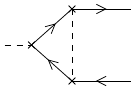}}
  \hfill
  \subfigure[]{\includegraphics[width=0.22\textwidth]{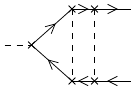}}
  \hfill
  \subfigure[]{\includegraphics[width=0.22\textwidth]{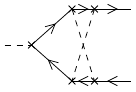}}
  \hfill
  \subfigure[]{\includegraphics[width=0.22\textwidth]{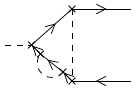}} \\[0.8em]
  \subfigure[]{\includegraphics[width=0.22\textwidth]{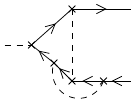}}
  \hfill
  \subfigure[]{\includegraphics[width=0.22\textwidth]{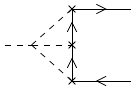}}
  \hfill
  \subfigure[]{\includegraphics[width=0.22\textwidth]{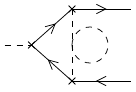}}
  \hfill
  \subfigure[]{\includegraphics[width=0.22\textwidth]{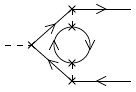}}
  \caption{The correction to Yukawa coupling strength $g$. The arrowed line denotes a fermion propagator, and the dashed line represents a boson propagator. The vertex with `$\times$' denotes the Yukawa interaction on the boundary.}
  \label{fig:yukawa_feynman}
\end{figure}

We outline the renormalization group (RG) analysis used to obtain the fixed point and critical exponents in the main text. 
We present the effective action, propagators, the RG flow equation, and the anomalous dimension, which follows the standard technical steps of dimensional regularization and minimal subtraction. 

The low-energy effective field theory describing the special phase reads,
\begin{align}
    S=\int_{\mathcal M}{\rm d}^dx \left(\frac12 \left(\partial_\mu\phi\right)^2+\frac{\lambda}{4!}\phi^4\right)+\int_{\partial\mathcal{M}}{\rm d}^{d-1}x\left( \bar\psi_i\gamma^\mu\partial_\mu\psi_i+g\phi\bar\psi_i\psi_i+h\phi^2\right),
\end{align}
where, $\psi_i$ denotes the $N$ flavors of four-component Dirac fermions (equivalently, $4N$ flavors of two-component Majorana fermions) and $\phi$ is the order parameter that represents the TRB $s$-wave pairing superconductor. 
The field theory is located in the $d$-dimensional semi-infinite spacetime, $\mathcal M=\{x_\mu|x_2>0\}$, with the boundary being $\partial\mathcal M=\{x_\mu|x_2=0\}$. 
Here, $\gamma^\mu$ denotes the Gamma matrix $\{\gamma^\mu,\gamma^\nu\}=2\delta^{\mu\nu}$, ${\rm Tr}\left[\gamma^\mu\gamma^\nu\right]=4\delta^{\mu\nu}$, $\bar\psi_i=\psi_i^\dagger\gamma^0$. The $\partial_\mu$ denotes derivatives with respect to (imaginary) times $x_0$ and spatial coordinates $x_i$, $i=1,2,\dots,d-1$. The kinetic term is $\gamma^\mu\partial_\mu\equiv\gamma^0\partial_0+\gamma^1\partial_1+\gamma^3\partial_3+\dots+\gamma^{d-1}\partial_{d-1}$ since $x_2=0$ for the boundary. 
The parameters $\lambda$ and $g$ represent the quartic boson self-interaction in the bulk and the fermion-boson Yukawa coupling at the boundary,
respectively. 
The bosonic propagator is obtained from the Gaussian part of the action, with the boundary condition, $\partial_y\phi|_{y=0}=2h\phi|_{y=0}$. The resulting boson propagator is given by,
\begin{align}
    D(p,y,y')=\frac{1}{2|p|}\left(e^{-|p||y-y'|}+\frac{|p|-2h}{|p|+2h}e^{-|p|(y+y')}\right),
\end{align}
where $|p|=\sqrt{\sum_{i\neq2}p_i^2}$ is the modulus of the $\left(d-1\right)$-dimensional momentum.
In particular, for the edge boson, $y=y'=0$, and at the special transition ($h=0$), the boson propagator reduces to,
\begin{align}
    D(p,0,0)=\frac{1}{|p|}\,.
\end{align}
The boundary fermion propagator takes the standard form,
\begin{align}
    G(k)=i\frac{\slashed{k}}{k^2}\,,
\end{align}
where $\slashed{k}=\sum_{\mu\neq2}k_\mu\gamma^\mu$ in $d-1$-dimension.

\begin{figure}[t]
  \centering
  \subfigure[]{\includegraphics[width=0.19\textwidth]{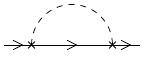}}
  \hfill
  \subfigure[]{\includegraphics[width=0.19\textwidth]{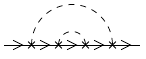}}
  \hfill
  \subfigure[]{\includegraphics[width=0.19\textwidth]{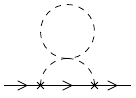}}
  \hfill
  \subfigure[]{\includegraphics[width=0.19\textwidth]{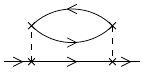}}
  \hfill
  \subfigure[]{\includegraphics[width=0.19\textwidth]{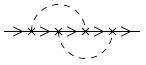}}
  \caption{Fermion self-energy diagrams.}
  \label{fig:fermion_self_ene}
\end{figure}

At the special transition point, whereas the bulk and boundary transition occur at the same time, we fix the bulk at the Wilson-Fisher fixed point,
\begin{align}
    \lambda^*=\frac{16\pi^2}{3}\epsilon+\frac{2448\pi^2}{1331}\epsilon^2\,.
\end{align}
To perform the renormalization group (RG) analysis, we introduce the renormalization factors of the wave function $\psi$, Yukawa coupling $g$, and the spin operators $S^i=\bar\psi\Gamma^i\psi$,
\begin{align}
    \psi=\sqrt{Z_{\psi}}\psi_{R}\,,\qquad g=\mu^{\frac{\epsilon}{2}}Z_{g}g_R\,,\qquad S^i=Z_{S}^iS_{R}^i\,,
\end{align}
where $\psi_R$, $g_R$ and $S_{R}^i$ denote the renormalized quantities.
The boundary boson will not be renormalized by the Dirac fermion, as one can show that the Feynman diagram involving fermions does not lead to pole in $\varepsilon$. 
The RG equation of the Yukawa coupling at the two-loop order reads
\begin{align}
    \frac{{\rm d}g}{{\rm d}\log\mu}=-\frac\epsilon2 g +f_0g^3+f_1g\lambda+f_2g^5 +f_3g^3\lambda+f_4g\lambda^2 \,,
\end{align}
with the coefficients
\begin{align}
    f_0&=\frac{2}{3\pi^2}\, ,\\    
    f_1&=-\frac{1}{32\pi^2}\,,\nonumber\\
    f_2&=-\frac{12N\pi^2+45-16\gamma_{\text{E}}-22\log2+16\log\pi}{72\pi^4}\,,\nonumber\\
    f_3&=\frac{-16-45\gamma_{\text{E}}+4\pi^2+72\log2+45\log\pi}{576\pi^4}\,,\nonumber\\
    f_4&=\frac{1}{256\pi^4}\,, \nonumber
\end{align}
where $\gamma_{\text{E}}$ denotes the Euler-Mascheroni constant. 
To determine the fixed point, we solve the beta equation $\beta(g^*)|_{\lambda=\lambda^*}=\frac{{\rm d}g}{{\rm d}\log\mu}\big|_{\lambda=\lambda^*}=0$. 
By setting $\epsilon=4-d=1$ and $N=\frac14$, we obtain $g^{*2}\approx3.258$. The nontrivial fixed point, $\{\lambda^*,g^{2*}\}$, marks the boundary Gross-Neveu-Yukawa (GNY) fixed point. 
The relevant Feynman diagrams are shown in Fig.~\ref{fig:yukawa_feynman}. %

The boundary Majorana fermion acquires an anomalous dimension:
\begin{align}
    \eta_{\psi}=\frac{g^2}{12\pi^2}+\frac{g^2\lambda\left(32-9\gamma+9\log\pi\right)}{1152\pi^4} -\frac{g^4\left(6N\pi^2+42+16\gamma_{\text{E}}-140\log2-16\log\pi\right)}{288\pi^4}\,.
\end{align}
At the boundary GNY fixed point, this yields $\eta_\psi\approx0.122$, leading to the boundary scaling dimension, $\Delta_\psi=\frac{d-2}{2}+\eta_\psi\approx0.622$.

\begin{figure}[t]
  \centering
  \hspace{0.2\textwidth} %
  \subfigure[]{\includegraphics[width=0.1\textwidth]{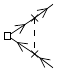}}
  \hspace{0.05\textwidth} %
  \subfigure[]{\includegraphics[width=0.1\textwidth]{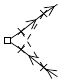}}
  \hspace{0.2\textwidth} %
  \caption{Feynman diagrams relevant to the boundary spin. The ``$\square$'' denotes the corresponding vertex $\Gamma^i$.}
  \label{fig:spin_Feynman}
\end{figure}

To obtain the boundary spin anomalous dimension, as $\eta_{S}^i=\frac{{\rm d}Z_{S}^i}{{\rm d}\log\mu}$, we consider the vertices $i\gamma^5$ and $i\gamma^1$, corresponding to $S^y$ and $S^z$, respectively. 
Note that the spin operator renormalization can be decomposed into fermion bilinear and vertex parts: $Z_{S}=Z_{\Gamma}^iZ_{\psi}$~\cite{rantner2002spin}. 
The vertex renormalization can be carried out by considering the diagram shown in Fig.~\ref{fig:spin_Feynman}(a), and the final results are provided in the main text.

This method can be systematically extended to higher orders. 
Here, we demonstrate the calculation of Fig.~\ref{fig:spin_Feynman}(b), which contains the most general construction. 
For vertex $i\gamma^1$, the corresponding diagram has $\gamma^1$ component
\begin{align}
    \frac{1}{4}\int\frac{{\rm d}^{d-1}p_1}{\left(2\pi\right)^2}\frac{{\rm d}^{d-1}p_2}{\left(2\pi\right)^2}{\rm Tr}\left[\gamma^1\frac{i\slashed{p}_2}{\left(p_2\right)^2}\frac{i\slashed{p}_1}{\left(p_1\right)^2}i\gamma^1\frac{i\slashed{p}_1}{\left(p_1\right)^2}\frac{i\big(\slashed{p}_1-\slashed{p}_2\big)}{\left(p_1-p_2\right)^2}\right]\frac{1}{\left|p_2\right|\left|p_1-p_2\right|}\,.
    \label{eq:spin_trace}
\end{align}
The trace produces the mixture of momenta. 
To address this problem, one can define the integrals,
\begin{align}
    S_0=&\int {\rm d}^{d-1}p_1{\rm d}^{d-1}p_2\ p_1^2\left(p_1\cdot p_2\right)F\left(\left|p_1\right|,\left|p_2\right|,p_1\cdot p_2\right)\,,\nonumber\\
    S_1=&\int {\rm d}^{d-1}p_1{\rm d}^{d-1}p_2\  p_1^2p_2^2F\left(\left|p_1\right|,\left|p_2\right|,p_1\cdot p_2\right)\,,\nonumber\\
    S_2=&\int {\rm d}^{d-1}p_1{\rm d}^{d-1}p_2\ \left(p_1\cdot p_2\right)^2F\left(\left|p_1\right|,\left|p_2\right|,p_1\cdot p_2\right)\,,
    \label{eq:spin_int}
\end{align}
with $F\left(\left|p_1\right|,\left|p_2\right|,p_1\cdot p_2\right)=\frac{1}{\left|p_1\right|^4\left|p_2\right|^5\left|p_1-p_2\right|}$. 
Due to rotation invariance, Eq.~\eqref{eq:spin_trace} can be summarized into the tensor forms,
\begin{align}
    T_{ijkl}&=\int {\rm d}^{d-1}p_1{\rm d}^{d-1}p_2\ p_{1i}p_{1j}p_{2k}p_{2l}F\left(\left|p_1\right|,\left|p_2\right|,p_1\cdot p_2\right)=A\delta_{ij}\delta_{kl}+B\left(\delta_{ik}\delta_{jl}+\delta_{il}\delta_{jk}\right)\,,\nonumber\\
    U_{ijkl}&=\int {\rm d}^{d-1}p_1{\rm d}^{d-1}p_2\ p_{1i}p_{1j}p_{1k}p_{2l}F\left(\left|p_1\right|,\left|p_2\right|,p_1\cdot p_2\right)=C\left(\delta_{ij}\delta_{kl}+\delta_{ik}\delta_{jl}+\delta_{il}\delta_{jk}\right)\,.
\end{align}
Contracting the tensor yields the following results:
\begin{align}
    A=\frac{\left(1+\text{dim}\right)S_1-2S_2}{\text{dim}\left(\text{dim}^2+\text{dim}-2\right)},\qquad B=\frac{\text{dim}S_2-S_1}{\text{dim}\left(\text{dim}^2+\text{dim}-2\right)},\qquad C=\frac{S_0}{\text{dim}^2+2\text{dim}}\,,
\end{align}
where ${\rm dim}=d-1$.

\end{document}